 \documentclass{JHEP3}
\usepackage{psfrag}
\usepackage{graphics}
\usepackage{amssymb,epsfig,amsmath,euscript,array,amsfonts,latexsym,mathrsfs,graphicx}
\newcounter{multieqs}

\newcommand{\be}{\begin{equation}}
\newcommand{\ee}{\end{equation}}

\newcommand{\bm}[1]{\mbox{\boldmath $#1$}}

\def\bd{\begin{document}}
\def\ed{\end{document}}
\def\nn{\nonumber}
\def\bea{\begin{eqnarray}}
\def\eea{\end{eqnarray}}
\let\bm=\bibitem
\let\la=\label

\def\npb#1#2#3{Nucl. Phys. {\bf{B#1}} #3 (#2)}
\def\plb#1#2#3{Phys. Lett. {\bf{#1B}} #3 (#2)}
\def\prl#1#2#3{Phys. Rev. Lett. {\bf{#1}} #3 (#2)}
\def\prd#1#2#3{Phys. Rev. {D \bf{#1}} #3 (#2)}
\def\cmp#1#2#3{Comm. Math. Phys. {\bf{#1}} #3 (#2)}
\def\cqg#1#2#3{Class. Quantum Grav. {\bf{#1}} #3 (#2)}
\def\nppsa#1#2#3{Nucl. Phys. B (Proc. Suppl.) {\bf{#1A}}#3 (#2)}
\def\ap#1#2#3{Ann. of Phys. {\bf{#1}} #3 (#2)}
\def\ijmp#1#2#3{Int. J. Mod. Phys. {\bf{A#1}} #3 (#2)}
\def\rmp#1#2#3{Rev. Mod. Phys. {\bf{#1}} #3 (#2)}
\def\mpla#1#2#3{Mod. Phys. Lett. {\bf A#1} #3 (#2)}
\def\jhep#1#2#3{J. High Energy Phys. {\bf #1} #3 (#2)}
\def\atmp#1#2#3{Adv. Theor. Math. Phys. {\bf #1} #3 (#2)}

\newcommand{\EQ}[1]{\begin{equation} #1 \end{equation}}
\newcommand{\AL}[1]{\begin{subequations}\begin{align} #1 \end{align}\end{subequations}}
\newcommand{\SP}[1]{\begin{equation}\begin{split} #1 \end{split}\end{equation}}
\newcommand{\ALAT}[2]{\begin{subequations}\begin{alignat}{#1} #2 \end{alignat}\end{subequations}}
\def\beqa{\begin{eqnarray}}
\def\eeqa{\end{eqnarray}}
\def\beq{\begin{equation}}
\def\eeq{\end{equation}}

\def\N{{\cal N}}
\def\sst{\scriptscriptstyle}
\def\thetabar{\bar\theta}
\def\Tr{{\rm Tr}}
\def\one{\mbox{1 \kern-.59em {\rm l}}}
 \def\Nh{\hat{N}}

\def\a{\alpha}      \def\da{{\dot\alpha}}
\def\b{\beta}       \def\db{{\dot\beta}}
\def\c{\gamma}  \def\G{\Gamma}  \def\cdt{\dot\gamma}
\def\d{\delta}  \def\D{\Delta}  \def\ddt{\dot\delta}
\def\e{\epsilon}        \def\vare{\varepsilon}
\def\f{\phi}    \def\F{\Phi}    \def\vvf{\f}
\def\h{\eta}
\def\k{\kappa}
\def\l{\lambda} \def\L{\Lambda}
\def\m{\mu} \def\n{\nu}
\def\o{\omega}
\def\p{\pi} \def\P{\Pi}
\def\r{\rho}
\def\s{\sigma}  \def\S{\Sigma}
\def\t{\tau}
\def\th{\theta} \def\Th{\Theta} \def\vth{\vartheta}
\def\X{\Xeta}
\def\z{\zeta}

\def\cA{{\cal A}} \def\cB{{\cal B}} \def\cC{{\cal C}}
\def\cD{{\cal D}} \def\cE{{\cal E}} \def\cF{{\cal F}}
\def\cG{{\cal G}} \def\cH{{\cal H}} \def\cI{{\cal I}}
\def\cJ{{\cal J}} \def\cK{{\cal K}} \def\cL{{\cal L}}
\def\cM{{\cal M}} \def\cN{{\cal N}} \def\cO{{\cal O}}
\def\cP{{\cal P}} \def\cQ{{\cal Q}} \def\cR{{\cal R}}
\def\cS{{\cal S}} \def\cT{{\cal T}} \def\cU{{\cal U}}
\def\cV{{\cal V}} \def\cW{{\cal W}} \def\cX{{\cal X}}
\def\cY{{\cal Y}} \def\cZ{{\cal Z}}

\def\ua{\underline{\alpha}}
\def\ub{\underline{\phantom{\alpha}}\!\!\!\beta}
\def\uc{\underline{\phantom{\alpha}}\!\!\!\gamma}
\def\um{\underline{\mu}}
\def\ud{\underline\delta}
\def\ue{\underline\epsilon}
\def\una{\underline a}\def\unA{\underline A}
\def\unb{\underline b}\def\unB{\underline B}
\def\unc{\underline c}\def\unC{\underline C}
\def\und{\underline d}\def\unD{\underline D}
\def\une{\underline e}\def\unE{\underline E}
\def\unf{\underline{\phantom{e}}\!\!\!\! f}\def\unF{\underline F}
\def\unm{\underline m}\def\unM{\underline M}
\def\unn{\underline n}\def\unN{\underline N}
\def\unp{\underline{\phantom{a}}\!\!\! p}\def\unP{\underline P}
\def\unq{\underline{\phantom{a}}\!\!\! q}
\def\unQ{\underline{\phantom{A}}\!\!\!\! Q}
\def\unH{\underline{H}}

\def\As {{A \hspace{-6.4pt} \slash}\;}
\def\bs {{b \hspace{-6.4pt} \slash}\;}
\def\Ds {{D \hspace{-6.4pt} \slash}\;}
\def\ds {{\del \hspace{-6.4pt} \slash}\;}
\def\ss {{\s \hspace{-6.4pt} \slash}\;}
\def\ks {{ k \hspace{-6.4pt} \slash}\;}
\def\ps {{p \hspace{-6.4pt} \slash}\;}
\def\pas {{{p_1} \hspace{-6.4pt} \slash}\;}
\def\pbs {{{p_2} \hspace{-6.4pt} \slash}\;}

\def\Fh{\hat{F}}
\def\Vh{\hat{V}}
\def\Xh{\hat{X}}
\def\ah{\hat{a}}
\def\xh{\hat{x}}
\def\yh{\hat{y}}
\def\ph{\hat{p}}
\def\xih{\hat{\xi}}

\def\psit{\tilde{\psi}}
\def\Psit{\tilde{\Psi}}
\def\tht{\tilde{\th}}

\def\At{\tilde{A}}
\def\Qt{\tilde{Q}}
\def\Rt{\tilde{R}}
\def\Nt{\tilde{N}}

\def\at{\tilde{a}}
\def\st{\tilde{s}}
\def\ft{\tilde{f}}
\def\pt{\tilde{p}}
\def\qt{\tilde{q}}
\def\vt{\tilde{v}}
\def\nt{\tilde{n}}

\def\delb{\bar{\partial}}
\def\bz{\bar{z}}
\def\bD{\bar{D}}
\def\bB{\bar{B}}

\def\bk{{\bf k}}
\def\bl{{\bf l}}
\def\bp{{\bf p}}
\def\bq{{\bf q}}
\def\br{{\bf r}}
\def\bx{{\bf x}}
\def\by{{\bf y}}
\def\bR{{\bf R}}
\def\bV{{\bf V}}

\def\su21{SU$(2|1)$}
\def\SU23{SU$(2|3)$}
\def\sp{{}}
\def\N2{${\cal N}=2$}
\def\4N{${\cal N}=4$}
\def\1N{${\cal N}=1$}
\def\half{\frac{1}{2}}
\def\d{\delta}\def\D{\Delta}\def\ddt{\dot\delta}

\def\pa{\partial} \def\del{\partial}
\def\xx{\times}
\def\uno{\mbox{1 \kern-.59em {\rm l}}}
\def\RE{\mbox{R \kern-1.19em {\rm I}}}
\def\IM{\mbox{I \kern-.79em {\rm I}}}
\def\betaR{\beta_{\sst R}}
\def\hf{{\sst \frac{1}{2}}}

\def\trp{^{\top}}
\def\inv{^{-1}}
\def\dag{{^{\dagger}}}
\def\pr{^{\prime}}

\def\rar{\rightarrow}
\def\lar{\leftarrow}
\def\lrar{\leftrightarrow}

\newcommand{\0}{\,\!}      %this is just NOTHING!
\def\one{1\!\!1\,\,}
\def\im{\imath}
\def\jm{\jmath}

\newcommand{\tr}{\mbox{tr}}
\newcommand{\slsh}[1]{/ \!\!\!\! #1}

\def\vac{|0\rangle}
\def\lvac{\langle 0|}

\def\hlf{\frac{1}{2}}
\def\ove#1{\frac{1}{#1}}

\def\Box{\square}
\def\ZZ{\mathbb{Z}}
\def\CC#1{({\bf #1})}
\def\bcomment#1{}
\def\bfhat#1{{\bf \hat{#1}}}
\def\VEV#1{\left\langle #1\right\rangle}

\newcommand{\ex}[1]{{\rm e}^{#1}} \def\ii{{\rm i}}

\def\rr{{\rm r}} \def\rs{{\rm s}}\def\rv{{\rm v}}
\def\ri{{\rm i}}\def\rj{{\rm j}}

\newcommand{\lrbrk}[1]{\left(#1\right)}
\newcommand{\sfrac}[2]{{\textstyle\frac{#1}{#2}}}

\font\mybb=msbm10 at 12pt
\def\bb#1{\hbox{\mybb#1}}

\font\myBB=msbm10 at 18pt
\def\BB#1{\hbox{\myBB#1}}

\title{Yangians in Deformed Super Yang-Mills Theories}

\author{Jay N. Ihry\\University of North Carolina, Chapel Hill, NC 27599-3255 \\ \email{jihry@physics.unc.edu}}

\abstract{
We discuss the integrability structure of deformed, four-dimensional \4N super Yang-Mills theories using Yangians.  We employ a recent procedure by Beisert and Roiban that generalizes the beta deformation of Lunin and Maldacena to produce \1N superconformal gauge theories, which have the superalgebra SU$(2,2|1)\times$U$(1)^2$.  The deformed theories, including those with the more general twist, were shown to have retained their integrable structure.  Here we examine the Yangian algebra of these deformed theories.  In a five field subsector, we compute the two cases of SU(2)$\times$U(1)$^3$ and \su21$\times$U(1)$^2$ as residual symmetries of SU$(2,2|1)\times$U(1)$^2$.  We compute a twisted coproduct for these theories, and show that only for the residual symmetry do we retain the standard coproduct.  The twisted coproduct thus provides a method for symmetry breaking.  However, the full Yangian structure of \SU23 is manifest in our subsector, albeit with twisted coproducts,  and provides for the integrability of the theory.  }
\keywords{Duality in Gauge Field Theories, AdS-CFT Correspondence, Extended Supersymmetry, Integrable Field Theories}
\preprint{arXiv:0802.3644}

\begin{document}

%%%%%%%%%%%%%%%%%%%%%%%%%%%%%
%%
%% SECTION: INTRODUCTION
%%
%%%%%%%%%%%%%%%%%%%%%%%%%%%%%
\begin{section}{Introduction}
\par With the advent of the of AdS/CFT correspondence, there has been much interest in conformal field theories.  In the mid 1990's \1N conformal field theories were constructed by exactly marginal deformations \cite{Leigh:1995ep}.  These \1N theories have the same particle content as the original \4N theory.  The \4N super Yang-Mills theory was broken to a \1N superconformal theory by the addition of the classical marginal deformation with the superpotential
\begin{equation}\label{margDef}
{\mathcal W} = i h \Tr \left( e^{i\pi\beta}\Phi_1\Phi_2\Phi_3 - e^{-i\pi\beta}\Phi_1\Phi_3\Phi_2 \right) + \frac{i h'}{3} \Tr \left( \Phi_1^3+\Phi_2^3+\Phi_3^3 \right).
\end{equation}
The condition for an exact marginal deformation, is that the parameters must obey 
\begin{equation}\label{margDefRed}
|h|^2\left( 1+ \frac{1}{N^2} \left( q-\bar{q}\right)^2 \right)+|h'|^2\frac{N^2-4}{2N^2} = g^2,
\end{equation}
where $h$,  $h'$, and $q=e^{i\pi \beta}$ are the deformation parameters and $g$ is the Yang-Mills coupling constant.  The large $N$ limit simplifies this condition, and if we set $h'=0$ as is commonly done, then $h=g$ is the requirement for a marginal deformation.  We will consider real $\beta$.
\par Lunin and Maldacena incorporated this deformation via a star product \cite{Lunin:2005jy} to find an \1N superconformal Yang-Mills theory with global U(1) symmetries.  They found the gravity dual of this theory through the AdS/CFT correspondence.  A three-parameter family of parameters replacing $\beta$ was given \cite{Frolov:2005dj}.  Using Bethe ansatz techniques, it was found in \cite{Roiban:2003dw}-\cite{Berenstein:2004ys} that the one-loop corrections in the large N limit of these deformed theories still provided an integrable spin chain Hamiltonian.  More general deformed integrable theories were provided in \cite{Beisert:2005if}.  These correspond to multi-parameter, \1N superconformal theories with the Lagrangian 
\be\label{defLag}\begin{array}{ccl}
{\mathcal L}&=& \frac{1}{g^2} \Tr \left( \frac{1}{4}F^{\mu\nu}F_{\mu\nu} + \left( D^{\mu} \bar{\Phi}^i  \right)\left(D_{\mu}\Phi_i\right)-\frac{1}{2}[\Phi_i,\Phi_j]_{C_{ij}}[\bar{\Phi}^i,\bar{\Phi}^j]_{C_{ij}}+\frac{1}{4}[\Phi_i,\bar{\Phi}^i][\Phi_j,\bar{\Phi}^j] \right. \\[2mm]
& & \left. \lambda_A \sigma^{\mu} D_{\mu} \bar{\lambda}^A -i([\lambda_4,\lambda_i]_{B_{4i}}\bar{\Phi}^i+[\bar{\lambda}^4,\bar{\lambda}^i]_{B_{4i}}\Phi_i) + \frac{i}{2}(\epsilon^{ijk}[\lambda_i,\lambda_j]_{B_{ij}}\Phi_k+\epsilon_{ijk}[\bar{\lambda}^i,\bar{\lambda}^j]_{B_{ij}}\bar{\Phi}^k)
\right),
\end{array}\ee
where $B_{ij}$ and $C_{ij}$ are related, and describe the deformations.  The gauge group is SU(N).  For these theories, amplitudes and the finiteness properties have been calculated \cite{Khoze:2005nd}-\cite{Kazakov:2007dy}.  Some further connections between integrability and deformed theories have been discussed in \cite{Bundzik:2005zg}-\cite{Beisert:2007jv}.
\par In this paper, we compute the Yangian structure for such deformed theories.  The ordinary symmetries are the \1N superconformal algebra SU($2,2|1$) with two global U(1) symmetries.  Following \cite{Beisert:2005if},\cite{Beisert:2003jj},\cite{Beisert:2003ys}, we will consider the one-loop dilatation operator, which we call the Hamiltonian, in a subsector of the full deformed theory.  This subsector has five one-particle states in the fundamental representation of \SU23.  In section 2 we discuss the algebraic structure of \SU23 and its Yangian extension.  In section 3 we give the two-site Hamiltonian as a quadratic Casimir and discuss the \SU23 Yangian symmetry of the undeformed theory \cite{Zwiebel:2006cb}-\cite{Dolan:2004ys}.  In section 4 we give the Hamiltonian for the deformed theory in this five field subsector, in the planar limit, and compute the Yangian generators for various cases including the Lunin-Maldacena deformation.  In section 5, we compute the twisted coproducts associated with multiparameter deformations \cite{Reshetikhin:1990ep,BasuMallick:1994pc}.  This structure was hinted at in \cite{Beisert:2005if}.  We show the residual symmetry of the deformed theory continues to use the standard coproduct while the remaining structure does not.  We illustrate this in two examples finding residual SU(2)$\times$U(1)$^3$ and \su21$\times$U(1)$^2$ symmetry, and discuss how the twisted coproduct is responsible for the smaller symmetry group for the deformed conformal gauge field theory.
\end{section}%INTRODUCTION

%%%%%%%%%%%%%%%%%%%%%%%%%%%%%
%%
%% SECTION: ALGEBRA
%%
%%%%%%%%%%%%%%%%%%%%%%%%%%%%%
\begin{section}{The \SU23 Sector}
\par We review the fields of this closed subsector and discuss its symmetry algebra.  The five fields include two complex fermions and three complex bosons, ${\mathbf \Phi}_I = \{\psi_1,\psi_2;\phi_1,\phi_2,\phi_3\}$.  We can express these fields as single particle states,
\be
\phi_a(i)|0\rangle = c^{\dag}_a(i) c^{\dag}_4(i)|0\rangle, \hspace{5mm} \psi_{\alpha}(j)|0\rangle = a^{\dag}_{\alpha} (j) c^{\dag}_4(j)|0\rangle,
\ee
where $1 \leq \alpha,\beta \leq 2$ and $1 \leq a,b \leq 3$ unless otherwise stated.  Site indices $i,j$ run over the length of the chain, $1\leq i,j \leq L$.  The oscillator (field), $c^{\dag}_4(i)$, is a remnant of the full PSU($2,2|4$) theory \cite{Beisert:2003jj} and is included to ensure the fermionic and bosonic properties of this oscillator representation.  The oscillator commutation relations are
\begin{equation}\label{osc defs2}
  [a^{\alpha}(i),a_{\beta}^{\dag}(j)] = \delta^{\alpha}_{\beta}\delta^i_j, \hspace{5mm}
  \{ c^a(i), c^{\dag}_b(j) \} = \delta^a_b\delta^i_j .
\end{equation}
The twenty-four generators of the \SU23 superalgebra have the explicit representation, at tree level ($g=0$),
\be
\begin{array}{c}
R^a\sp_b = c^{\dag}_b c^a - \frac{1}{3} \delta^a_b c^{\dag}_c c^c, \hspace{5mm} L^\alpha\sp_{\beta} = a^{\dag}_\beta a^{\alpha} - \frac{1}{2} \delta^{\alpha}_{\beta}a^{\dag}_{\gamma}a^{\gamma}, \\[2mm]
D =  c^{\dag}_c c^c + \frac{3}{2} a^{\dag}_{\gamma}a^{\gamma}, \hspace{5mm}S^{\gamma}\sp_{c} = c^{\dag}_c a^{\gamma}, \hspace{5mm} Q^c\sp_{\gamma} = a^{\dag}_{\gamma} c^c.
\end{array}\ee
%%%
% SUBSECTION: SU23
%%%
\vskip10pt
\noindent{\em The SU$(2|3)$ Algebra}
\par A single index basis for the symmetry generators of the ordinary \SU23 algebra is given in Appendix A.  The symmetry generators for SU(3) and SU(2) carry indices $\{1,\dots , 8\}$ and $\{9,10,11\}$, respectively; the dilation generator has index 12, and the odd generators are labeled by $\{13, \dots, 24\}$.  A detailed analysis of the \SU23 algebra and resulting spin chain can be found in \cite{Beisert:2003jj,Beisert:2003ys}.  The symmetry generators close the algebra
\be\label{comm}
\left[ J^A , J^B \right\} = f^{AB}\sp_{C} J^C = f^{ABD}g_{DC} J^C,
\ee
where the structure constants $f^{ABC}$ and the metric $g_{AB}$ of \SU23 can also be found in Appendix A.  This basis allows for a simple presentation of the Yangian defining relations.  
%SU23
\vskip20pt
%%%
% SUBSECTION SU(2|3) YANGIAN
%%%
\noindent{{\em The \SU23 Yangian Algebra}}
\par An infinite-dimensional extension of the \SU23 algebra, called the Yangian \cite{Drinfeld:1985rx,Drinfeld:1986in}, has a tree level representation in terms of the ordinary generators 
\begin{equation}\label{yangDef}
Q^A_0 = -f^A\sp_{CB} \sum_{i<j} J^B_0(i) J^C_0(j).
\end{equation}
\par This representation takes into account the superalgebra properties of the Lie algebra \cite{Zwiebel:2006cb}.   The super Yangian algebra defining relations are 
\be\label{alg} \left[ J^A , J^B \right\} = f^{AB}\sp_{C}J^C, \ee
\be \left[ J^A , Q^B \right\} = f^{AB}\sp_{C} Q^C,\ee
\be\label{serre} \left[ Q^{\left[ A \right.}, \left[ Q^{B} , J^{\left. C \right]} \right\} \right\} 
= \alpha f^{AG}\sp_{D}f^{BH}\sp_{E}f^{CK}\sp_{F}f_{GHK}J^{\left\{D\right.}J^EJ^{\left. F \right\}}. \ee
The last is the Serre relation which holds because the generators $J^A$ are in a certain representation.  The constant $\alpha$ depends on the choice of basis.  Here $J^{\left\{D\right.}J^EJ^{\left. F \right\} }$ is the totally symmetric product, with an additional minus sign for the exchange of two odd generators.
%SU23 YANGIAN
\end{section}%THE ALGEBRA

%%%%%%%%%%%%%%%%%%%%%%%%%%%%%
%%
%% SECTION QUAD CASIMIR AND 2 SITE HAMILTONIAN
%%
%%%%%%%%%%%%%%%%%%%%%%%%%%%%%
\begin{section}{The Hamiltonian}
\par A useful feature of this sector is the relationship between the quadratic Casimir and the two-site Hamiltonian.  
%%%%
%% SUBSECTION TWO-SITE HAMILTONIAN
%%%%
%\begin{subsection}{Two-Site Hamiltonian}
A Hamiltonian, of generic length $L$, was found in \cite{Beisert:2003jj}\footnote{Length changing, for the \SU23 sector, is required for higher than one-loop order, but here we consider only one-loop, ${\cal O}(g^2)$.}.  The two-site Hamiltonian is
\begin{align}\label{oscHam}
 H_{12} =&\left(c^{\dag}_a(1)c^{\dag}_b(2)-c^{\dag}_b(1)c^{\dag}_a(2)\right)c^b(2)c^a(1)+ \left( c^{\dag}_a(1)a^{\dag}_{\alpha}(2)+a^{\dag}_{\alpha}(1)c^{\dag}_a(2)\right)a^{\alpha}(2)c^a(1)  \cr
 &+ \left( a^{\dag}_{\alpha}(1)c^{\dag}_a(2)+c^{\dag}_a(1)a^{\dag}_{\alpha}(2)\right)c^a(2)a^{\alpha}(1)+ \left( a^{\dag}_{\alpha}(1) a^{\dag}_{\beta}(2)+a^{\dag}_{\beta}(1)a^{\dag}_{\alpha}(2)\right)a^{\beta}(2)a^{\alpha}(1). \cr
\end{align} 
One can explicitly check, the Hamiltonian above has two eigenstates with eigenvalues 0 and 2.  These correspond to symmetric and antisymmetric two-particle states and are discussed below.
%\end{subsection}%two-site HAMILTONIAN

%%%%
%% SUBSECTION QUADRATIC CASIMIR
%%%%
%\begin{subsection}{Quadratic Casimir}
\par  The two-site quadratic Casimir is the operator, $g_{AB}J^AJ^B=g_{AB}(J(1)^A+J^A(2))(J^B(1)+J^B(2))$.  And can be explicitly shown,
\begin{equation}\label{quadCas}
 g_{AB}J^AJ^B = {\textstyle \frac{1}{3}}D^2+{\textstyle \frac{1}{2}} L^{\gamma}{}_{\delta}L^{\delta}{}_{\gamma}-{\textstyle \frac{1}{2}} R^c{}_dR^d{}_c -{\textstyle\frac{1}{2}} \left[ Q^c{}_{\gamma} , S^{\gamma}{}_c \right].
\end{equation}
The single site quadratic Casimir acting on any two-particle state $|\eta\rangle$, is zero. So, $g_{AB}J^A(1)J^B(1)|\eta\rangle = g_{AB}J^A(2)J^B(2)|\eta\rangle =0$ and the cross term piece is \break $2g_{AB}J^A(1)J^B(2)  = H_{12}$,
where $H_{12}$ is given in \eqref{oscHam}.  So the two-site Hamiltonian can be identified with the quadratic Casimir
\begin{equation}\label{equals}
 H_{12}|\eta\rangle = g_{AB}[J^A(1)+J^A(2)] [J^B(1)+J^B(2)]|\eta\rangle,
\end{equation}
when acting the states.
%\end{subsection}%QUADRATIC CASIMIR
%%%%
%% SUBSECTION THE TWO TOWERS
%%%%
%\begin{subsection}{The Two Towers}
\par For calculations with Yangians it is useful to use eigenstates of the Hamiltonian.  The two-particle eigenstates are symmetric or antisymmetric in the site indices.   We define them as, $|{\mathbf \Phi}_I {\mathbf \Phi}_J \rangle_{\pm} = |{\mathbf \Phi}_I(1){\mathbf \Phi}_J(2) \rangle \pm |{\mathbf \Phi}_I(2){\mathbf \Phi}_J(1)\rangle$.  The explicit representation of the symmetric states is,
\begin{equation}\label{symm}\begin{array}{ccl}
|ab\rangle_{+}                 &=& -\left( c^{\dag}_a(1)c^{\dag}_b(2) + c^{\dag}_b(1)c^{\dag}_a(2) \right)c^{\dag}_4(1)c^{\dag}_4(2) |0\rangle, \\
|a\beta\rangle_{+}          &=& \left(c^{\dag}_a(1)a^{\dag}_{\beta}(2) - a^{\dag}_{\beta}(1)c^{\dag}_a(2) \right)c^{\dag}_4(1)c^{\dag}_4(2) |0\rangle, \\
|\alpha\beta\rangle_{+} &=& \left(a^{\dag}_{\alpha}(1)a^{\dag}_{\beta}(2) - a^{\dag}_{\beta}(1)a^{\dag}_{\alpha}(2) \right)c^{\dag}_4(1)c^{\dag}_4(2) |0\rangle . 
\end{array}\end{equation}
The representation of the antisymmetric states is 
\begin{equation}\label{anti}\begin{array}{ccl}
|ab\rangle_{-}                 &=& -\left( c^{\dag}_a(1)c^{\dag}_b(2) - c^{\dag}_b(1)c^{\dag}_a(2) \right)c^{\dag}_4(1)c^{\dag}_4(2) |0\rangle, \\
|a\beta\rangle_{-}          &=& \left(c^{\dag}_a(1)a^{\dag}_{\beta}(2) + a^{\dag}_{\beta}(1)c^{\dag}_a(2) \right)c^{\dag}_4(1)c^{\dag}_4(2) |0\rangle, \\
|\alpha\beta\rangle_{-} &=& \left(a^{\dag}_{\alpha}(1)a^{\dag}_{\beta}(2) + a^{\dag}_{\beta}(1)a^{\dag}_{\alpha}(2) \right)c^{\dag}_4(1)c^{\dag}_4(2) |0\rangle,
\end{array}\end{equation}
These two groups, symmetric and antisymmetric, make up two towers.  Ordinary symmetry generators of \SU23, move one up and down in each tower, while Yangian generators move from one tower to a linear combination in the other. 
%%%
% SUBSECTION:  [H_2(1,2),Q^A] = q^A
%%%
%\begin{subsection}{$\left[H_2(1,2), Q^A\right] = q^A$}
\par Commutation of the dilatation operator with the other symmetry generators gives the anomolous dimension of that operator, $[D,J^A]=(\text{dim} J^A)J^A$. This relation holds for the Yangian, $[D,Q^A]=(\text{dim} J^A)Q^A$.  Assuming these relations hold to all order of the Yang-Mills coupling, as discussed in \cite{Dolan:2003uh}, we expand the operators, $[ D, Q^A ]  = [ D_0 + g^2_{\text{YM}}D_{2} + \cdots , Q_0^A + g_{\text{YM}}Q_1^A + g^2_{\text{YM}}Q_2+ \cdots] $ and group in powers of the Yang-Mills coupling $g$,
\be
 (\text{dim} J^A)(Q^A_0 + g_{\text{YM}}Q^A_1+g^2_{\text{YM}}Q^A_2) + g^2_{\text{YM}}[D_2,Q^A_0]\approx (\text{dim} J^A)Q^A.
\ee
To find that $[D_2,Q^A_0]$ must be zero.  In PSU($2,2|4$), an explicit check of the commutator gives the lattice derivative or `edge effects' of the system, $[D_2,Q^A_0] = q^A \sim 0$, $q_{1L}^A = J^A(1) - J^A(L)$.  Our Yangian in \SU23, while not a subalgebra of the Yangian of PSU$(2,2|4)$, maintains this relation.  To see this we introduce the identity,
\be\label{ident}
\left[ g_{AB}J(1)^AJ(2)^B , q_{12}^C \right] = 4Q_{12}^C,
\ee
where $Q^A_{12}$ is the two-site version of the bare generator $Q_0^A$.  So we have $\frac{1}{4}  \left[ H_{12} , q_{12}^A \right] = Q_{12}^A$, and the one-loop calculation becomes
\be
\left[H_{12} , Q_{12}^A\right] = {\textstyle \frac{1}{4}} \left( H_{12}^2 q_{12}^A - 2 H_{12}q_{12}^AH_{12}+q_{12}^AH_{12}^2 \right).
\ee
From the previous section, the Yangian acting on a tower moves it to the other tower (i.e. moves a symmetric state to an antisymmetric and vice versa).  From Eq. \eqref{ident}, if the Yangian produces this type of movement so must the edge effect, $q^A$.  Recalling the values under the Hamiltonian of the two-particle states $H_{12}|{\bf \Phi}_1{\bf \Phi}_2\rangle _{+} = 0, H_{12}|{\bf \Phi}_1{\bf \Phi}_2\rangle _{-} = 2 |{\bf \Phi}_1{\bf \Phi}_2\rangle _{-}$, we see the middle term vanishes under both symmetric and antisymmetric states.  We find that
\be
  \left[H_{12}, Q_{12}^A\right]|{\bf \Phi}_1{\bf \Phi}_2\rangle_{\pm} =  q_{12}^A |{\bf \Phi}_1{\bf \Phi}_2\rangle_{\pm},
\ee
which is the two-site version of the edge effect described above in the \SU23 sector.
%\end{subsection}% [H_2,Q]=q

\end{section} %QUAD CASIMIR AND two-particle HAM

%%%%%%%%%%%%%%%%%%%%%%%%%%%%%%%%%%%%
%%
%% SECTION: DEFORMED HAMILTONIAN
%%
%%%%%%%%%%%%%%%%%%%%%%%%%%%%%%%%%%%%
\begin{section}{The Deformed Hamiltonian }
\par We turn to the twist deformation found in \cite{Beisert:2005if}.  A second solution to the graded Yang-Baxter equation for the \SU23 sector was given 
\be\label{defR}
\tilde{R} = \frac{1}{u+i} \left( u e^{-iB_{ij}}{\mathcal I}^{kl}_{ij} + i{\mathcal P}^{kl}_{ij} \right).
\ee
This deformed R-matrix, is the conventional R-matrix solution to the Yang-Baxter equation with additional phases.   The identity and projection operators are ${\mathcal I}^{kl}_{ij} = \delta^k_i \delta^l_j$ and ${\mathcal P}^{kl}_{ij} = \delta^l_i\delta^k_j$.  The deformed monodromy matrix is defined
\begin{equation}
\tilde{{\mathcal T}}^{b;\beta_1\dots\beta_L}_{a;\alpha_1\dots\alpha_L} = \tilde{R}^{b_{L-1}\beta_{L}}_{a\alpha_L}\tilde{R}^{b_{L-2}\beta_{L-1}}_{b_{L-1}\alpha_{L-1}} \cdots \tilde{R}^{b_{1}\beta_{2}}_{b_{2}\alpha_{2}}\tilde{R}^{a \beta_{1}}_{b_{1}\alpha_{1}}
\exp \left[ i \pi \sum^{L}_{i=1} \sum^{i-1}_{j=1}([\alpha_i] + [\beta_i])[\alpha_j]\right],
\end{equation}
where the $Z_2$ graded set of states is denoted by the $[ \alpha_i ]$. Derived from the trace of the monodromy matrix, the deformed transfer matrix is $\tilde{{\mathcal T}}(u) = (-)^{[a]}\tilde{{\mathcal T}}^a_a (u)$.  Like the normal case, we find the deformed Hamiltonian is the logarithmic derivative of the deformed transfer matrix, $\tilde{{\cal H}} = \left. -i \left(\tilde{{\mathcal T}}(u^*)\right)^{-1}\frac{d}{du}\tilde{{\mathcal T}}(u)\right|_{u=u^*}$.  This is more closely examined for two-particle states in the following section.
%%%
% SUBSECTION:  DEFORMED two-site HAMILTONIAN
%%%
\begin{subsection}{Deformed Two-Site Hamiltonian}
\par The two-site transfer matrix is $\tilde{T}(u) = \tilde{R}^{b_1\beta_2}_{a\alpha_2}\tilde{R}^{a\beta_1}_{b_1\alpha_1}\exp \left[ i \pi ([\alpha_2]+[\beta_2])[\alpha_1] \right] $.  Using normal procedures for calculating the Hamiltonian, we find the logarithmic derivative of the deformed transfer matrix and expand at $u^* = 0$,
\begin{equation}
\tilde{{\mathcal H}} = \left( \delta^{\beta_1}_{\alpha_1}\delta^{\beta_2}_{\alpha_2} - \delta^{\beta_2}_{\alpha_1}\delta^{\beta_1}_{\alpha_2} e^{-iB_{\alpha_1\alpha_2}}\right) +
\left( \delta^{\beta_1}_{\alpha_1}\delta^{\beta_2}_{\alpha_2} - \delta^{\beta_2}_{\alpha_1}\delta^{\beta_1}_{\alpha_2} e^{-iB_{\alpha_2\alpha_1}}\right)
= \left( \tilde{{\mathcal H}}^{\beta_1\beta_2}_{\alpha_1\alpha_2}\right) +\left( \tilde{{\mathcal H}}^{\beta_2\beta_1}_{\alpha_2\alpha_1}\right).
\end{equation}
Define the deformed two-site Hamiltonians, $\tilde{H}_{12} \equiv \tilde{{\mathcal H}}^{\beta_1\beta_2}_{\alpha_1\alpha_2}$ and $\tilde{H}_{21} \equiv \tilde{{\mathcal H}}^{\beta_2\beta_1}_{\alpha_2\alpha_1}$.  Examining closer, we see a phase is obtained under interchange of two fields.  
\begin{equation}\label{defHam}\begin{array}{rcl}
 \tilde{H}_{12}& = &  \left(c^{\dag}_a(1)c^{\dag}_b(2)-e^{-iB_{ab}}c^{\dag}_b(1)c^{\dag}_a(2)\right)c^b(2)c^a(1) \\[2mm]
                           & & +  \left( c^{\dag}_a(1)a^{\dag}_{\alpha}(2)+e^{-iB_{a\alpha}}a^{\dag}_{\alpha}(1)c^{\dag}_a(2)\right)a^{\alpha}(2)c^a(1)  \\[2mm]
                           && +  \left( a^{\dag}_{\alpha}(1)c^{\dag}_a(2)+e^{-iB_{\alpha a}}c^{\dag}_a(1)a^{\dag}_{\alpha}(2)\right)c^a(2)a^{\alpha}(1) \\[2mm]
                          & & +  \left( a^{\dag}_{\alpha}(1) a^{\dag}_{\beta}(2)+e^{-iB_{\alpha\beta}}a^{\dag}_{\beta}(1)a^{\dag}_{\alpha}(2)\right)a^{\beta}(2)a^{\alpha}(1).
\end{array}\end{equation}
The antisymmetric matrix $B_{AB}$ is a matrix of phases formed from the charges of the Cartan generators of the original R symmetry SU(4) and is described in \cite{Beisert:2005if},
\be
{\mathbf B} = \left( 
  \begin{array}{cccccc}
    0&-\gamma_3&+\gamma_2&\frac{1}{2} \left(\gamma_2-\gamma_3\right)&\frac{1}{2}\left(\gamma_2-\gamma_3\right)\\
    +\gamma_3&0&-\gamma_1&\frac{1}{2}\left(\gamma_3-\gamma_1\right)&\frac{1}{2}\left(\gamma_3-\gamma_1\right)\\
    -\gamma_2&+\gamma_1&0&\frac{1}{2}\left(\gamma_1-\gamma_2\right)&\frac{1}{2}\left(\gamma_1-\gamma_2\right)\\
    \frac{1}{2}\left(\gamma_3-\gamma_2\right)&\frac{1}{2}\left(\gamma_1-\gamma_3\right)&\frac{1}{2}\left(\gamma_2-\gamma_1\right)&0&0\\
    \frac{1}{2}\left(\gamma_3-\gamma_2\right)&\frac{1}{2}\left(\gamma_1-\gamma_3\right)&\frac{1}{2}\left(\gamma_2-\gamma_1\right)&0&0
  \end{array}
\right).
\ee
The deformation parameters $\gamma_i$ are three real constants.  The eigenstates of the deformed Hamiltonian are  
\begin{equation}\label{defEigStates}\begin{array}{ccl}
\widetilde{|ab\rangle}_{\pm}                 &=& -\left(e^{iB_{ab}/2}c^{\dag}_a(1)c^{\dag}_b(2) \pm e^{-iB_{ab}/2}c^{\dag}_b(1)c^{\dag}_a(2) \right)c^{\dag}_4(1)c^{\dag}_4(2) |0\rangle, \\[3mm]
\widetilde{|a\beta\rangle}_{\pm}          &=& \left(e^{iB_{a\beta}/2}c^{\dag}_a(1)a^{\dag}_{\beta}(2) \mp e^{-iB_{a\beta}/2}a^{\dag}_{\beta}(1)c^{\dag}_a(2) \right)c^{\dag}_4(1)c^{\dag}_4(2) |0\rangle, \\[3mm]
\widetilde{|\alpha\beta\rangle}_{\pm} &=& \left(e^{iB_{\alpha\beta}/2}a^{\dag}_{\alpha}(1)a^{\dag}_{\beta}(2) \mp e^{-iB_{\alpha\beta}/2}a^{\dag}_{\beta}(1)a^{\dag}_{\alpha}(2) \right)c^{\dag}_4(1)c^{\dag}_4(2) |0\rangle . 
\end{array}\end{equation}
As before, they have eigenvalues $\tilde{H}_2 \widetilde{|+\rangle} = 0 \widetilde{|+\rangle}$ and $\tilde{H}_2\widetilde{|-\rangle} = 2 \widetilde{|-\rangle}$.  Note that special cases of repeated fields will never receive phase corrections.
\end{subsection}%DEFORMED two-site HAMILTONIAN

%%%
% SUBSECTION: GAMMA1=GAMMA2=GAMMA3
%%%
\begin{subsection}{Case 1: $\gamma_1=\gamma_2=\gamma_3$}
\par Since we are interested in deformed theories that have \1N superconformal symmetry, we first examine the case of phase deformations with all parameters equal, $\gamma_1=\gamma_2=\gamma_3=\gamma$.  The resultant nonzero phases are $B_{13}=B_{21}=B_{32} =\gamma$ and give a residual SU(2)$\times$U(1)$^3$ symmetry.  This is the beta deformation of Lunin and Maldacena \cite{Lunin:2005jy,Beisert:2005if}, but restricted to our five field subsector.  So the nonzero commutation relations after deformation are just the SU(2) algebra
\be
  \left[ L^{\alpha}\sp_{\beta} , L^{\gamma}\sp_{\delta} \right] = \delta^{\alpha}_{\delta} L^{\gamma}\sp_{\beta} - \delta^{\gamma}_{\beta} L^{\alpha}\sp_{\delta}.
\ee
The three U(1) generators are $U_1 = {\textstyle \frac{3}{4}}c^{\dag}_4c^4-{\textstyle \frac{1}{4}}c^{\dag}_c c^c$, $U_2 = c^{\dag}_2c^2-c^{\dag}_3c^3$, and $U_3 = c^{\dag}_2c^2-c^{\dag}_1c^1$.  Two-particle eigenstates of the deformed Hamiltonian have a phase on the states with two SU(3) fields and no phase any of the additional states.  
\begin{equation}\label{defEigStates1}\begin{array}{ccl}
\widetilde{|ab\rangle}_{\pm}                 &=& -\left(e^{iB_{ab}/2}c^{\dag}_a(1)c^{\dag}_b(2) \pm e^{-iB_{ab}/2}c^{\dag}_b(1)c^{\dag}_a(2) \right)c^{\dag}_4(1)c^{\dag}_4(2) |0\rangle, \\[3mm]
\widetilde{|a\beta\rangle}_{\pm}          &=& \left(c^{\dag}_a(1)a^{\dag}_{\beta}(2) \mp a^{\dag}_{\beta}(1)c^{\dag}_a(2) \right)c^{\dag}_4(1)c^{\dag}_4(2) |0\rangle, \\[3mm]
\widetilde{|\alpha\beta\rangle}_{\pm} &=& \left(a^{\dag}_{\alpha}(1)a^{\dag}_{\beta}(2) \mp a^{\dag}_{\beta}(1)a^{\dag}_{\alpha}(2) \right)c^{\dag}_4(1)c^{\dag}_4(2) |0\rangle. 
\end{array}\end{equation}
\par If we try to examine the one-loop quantity, $[\tilde{H} , Q^A_0 ]$, using the Yangian from the undeformed \SU23 theory, we would find
\be\left[\tilde{H}, Q^A\right] \widetilde{|ab\rangle}_{\pm} = q^A \widetilde{|ab\rangle}_{\pm},\hspace{3mm}
\left[\tilde{H}, Q^A\right] \widetilde{|a\alpha\rangle}_{\pm}=q^A \widetilde{|a\alpha\rangle}_{\pm}, \hspace{3mm}
\left[\tilde{H}, Q^A\right] \widetilde{|\alpha\beta\rangle}_{\pm}=q^A \widetilde{|\alpha\beta\rangle}_{\pm},
\ee
only for $A = \{ A | J^A \in SU(2)\times\text{U}(1)^3 \}$.
\vskip20pt
\end{subsection}%GAMMA1=GAMMA2=GAMMA3
%%%
% SUBSECTION: GAMMA1=GAMMA2=-GAMMA3 CASE
%%%
\begin{subsection}{Case 2: $\gamma_1=\gamma_2=-\gamma_3$}
\par Another \1N superconformal theory, embedded differently in the original PSU$(2,2|4)$ algebra, is given by $\gamma_1=\gamma_2=-\gamma_3$.  The nonzero elements of the antisymmetric matrix are $B_{ab}: B_{12}=B_{13}=-B_{23} = \gamma$ and $B_{a\alpha}: B_{1\alpha} = -B_{2\beta} = \gamma$.\footnote{For the remainder of this section, $1\leq a,b \leq 2$, $1\leq \alpha\beta \leq 2$.}  The residual symmetry is \su21$\times$U$(1)^2$.  This symmetry algebra has a richer structure containing a superalgebra containing $\{L^{\alpha}\sp_{\beta}, Q^3\sp_{\alpha},S^{\alpha}\sp_{3}, R\}$  and the two U(1)s: $R = a^{\dag}_{\gamma}a^{\gamma} + 2c^{\dag}_{c}c^{c}$, $U_2 = c^{\dag}_{2}c^{2} - c^{\dag}_{1}c^{1}$,  $U_3 = c^{\dag}_{4}c^{4} - c^{\dag}_{2}c^{2}$.  The nonzero commutation relations for this form of the embedding are
\be\label{SU21Comms}\begin{array}{c}
\left[L^{\alpha}\sp_{\beta},J_{\gamma} \right] = \delta^{\alpha}_{\gamma}J_{\beta}-\frac{1}{2} \delta^{\alpha}_{\beta}J_{\gamma}, \hspace{5mm} \left[L^{\alpha}\sp_{\beta},J^{\gamma} \right] = -\delta^{\gamma}_{\beta}J_{\alpha} + \frac{1}{2} \delta^{\alpha}_{\beta} J^{\gamma}, \\[3mm]
\left[ R, S^{\alpha}\sp_{3} \right] = S^{\alpha}\sp_{3}, \hspace{5mm} \left[ R, Q^{3}\sp_{\alpha} \right] = - Q^{3}\sp_{\alpha}, \hspace{5mm} \left\{S^{\alpha}\sp_{3},Q^{3}\sp_{\beta} \right\} = L^{\alpha}\sp_{\beta} + \frac{1}{2} \delta^{\alpha}_{\beta} R. 
\end{array}\ee
We could again try to compute with the tree level Yangian in the deformed theory, however we would find that unless we use $Q^A$ with $\{A \in \text{SU}(2|1)\times\text{U}(1)^2\}$ and restrict to eigenstates whose one particle fields lie in the fundamental representation of the residual symmetry, the standard form of the tree level Yangian \eqref{yangDef} is not useful.
\end{subsection}%GAMMA1=GAMMA2=-GAMMA3
\par Therefore, we look for the appropriate form of the tree level Yangian from the deformed transfer matrix.  In \cite{BasuMallick:1994pc}, a twisted R-matrix is derived via a Reshetikhin twist \cite{Reshetikhin:1990ep} leading to a deformed coproduct.  Our deformed R-matrix is a supersymmetric version of this, as briefly mentioned in \cite{Beisert:2005if}.  So we will use a twisted coproduct to compute the tree level Yangian.
\end{section}%DEFORMED HAMILTONIAN

%%%%%%%%%%%%%%%%%%%%%%%%%%%%%%%%%%%%
%%
%% SECTION TWISTED COPRODUCTS
%%
%%%%%%%%%%%%%%%%%%%%%%%%%%%%%%%%%%%%
\begin{section}{Twisted Coproducts}
\par We identify the deformed R-matrix in Eq. \eqref{defR} with a multiparameter form \cite{Reshetikhin:1990ep,BasuMallick:1994pc}.  This requires a twisted coproduct on our generators.  For an algebra ${\cal A}$, a coproduct is a homomorphic map $\Delta: {\cal A} \rightarrow {\cal A}$ which brings a single site representation into a double-site representation, a double-site into a triple-site, etc \cite{Drinfeld:1986in,Drinfeld:1985rx}. Here, ${\cal A}$ is the Yangian of \SU23.  We forego our single index notation because this coproduct is dependent on the specifics of our generators and is easier to use with double indices.\footnote{A brief description can be found in Appendix B.}  The twisted coproduct for the ordinary generators is
\be\label{copr}\begin{array}{ccl}
\Delta R^{a}\sp_{b} &=& K_{ab} \otimes R^{a}\sp_{b} + R^{a}\sp_{b}\otimes K_{ba},\\[2mm]
\Delta L^{\alpha}\sp_{\beta} &=& K_{\alpha\beta} \otimes L^{\alpha}\sp_{\beta} + L^{\alpha}\sp_{\beta} \otimes K_{\beta\alpha},\\[2mm]
\Delta Q^{c}\sp_{\gamma}&=& K_{c\gamma} \otimes Q^{c}\sp_{\gamma} + Q^{c}\sp_{\gamma}\otimes K_{\gamma c},\\[2mm]
\Delta S^{\gamma}\sp_{c}&=& K_{\gamma c} \otimes S^{\gamma}\sp_{c} + S^{\gamma}\sp_{c} \otimes K_{c\gamma}, \\[2mm]
\Delta D &=& 1 \otimes D + D \otimes 1.
\end{array}\ee
As before, $1 \leq a,b \leq 3$ and $1 \leq \alpha,\beta \leq 2$ and now $1 \leq I,J,K \leq 5$.  The twisted coproducts depend on the antisymmetric parameters, $\alpha_{IJ} = - \alpha_{JI}$, which reside in $K_{IJ} = \exp \left[ \frac{i}{2} \sum_{K=1}^{5}\left( \alpha_{IK} - \alpha_{JK} \right) E_{KK} \right]$.\footnote{Twisted coproducts can be generated from a deforming function, $F$, such that $\Delta^{(F)} = F\Delta^{(0)}F^{-1}$, where $\Delta^{(0)}$ is the standard coproduct \cite{Reshetikhin:1990ep, Drinfeld:1985rx}.  The standard coproduct corresponds to $K_{IJ}=1$.}  So,
\be\begin{array}{c}
K_{ab} = e^{\frac{i}{2}\left(\alpha_{a \gamma} - \alpha_{b\gamma} \right)E_{\gamma\gamma} + \frac{i}{2}\left(\alpha_{ac} - \alpha_{bc}\right)E_{cc}} = K^{-1}_{ba}, \\[2mm]
K_{\alpha\beta} = e^{\frac{i}{2}\left( \alpha_{\alpha\gamma} - \alpha_{\beta\gamma}\right)E_{\gamma\gamma} + \frac{i}{2}\left( \alpha_{\alpha c} - \alpha_{\beta c} \right)E_{cc}} = K^{-1}_{\beta\alpha}, \\[2mm]
K_{a \alpha} = e^{\frac{i}{2}\left(\alpha_{a\gamma} - \alpha_{\alpha\gamma} \right)E_{\gamma\gamma} + \frac{i}{2}\left(\alpha_{ac} - \alpha_{\alpha c} \right)E_{cc}} = K^{-1}_{\alpha a}.
\end{array}\ee
The quadratic Casimir of the \SU23 algebra with this twisted coproduct is
\be
 g_{AB}\Delta J^A \Delta J^B = {\textstyle \frac{1}{3}}\Delta D\Delta D+{\textstyle \frac{1}{2}} \Delta L^{\gamma}{}_{\delta}\Delta L^{\delta}{}_{\gamma}-{\textstyle \frac{1}{2}} \Delta R^c{}_d\Delta R^d{}_c -{\textstyle \frac{1}{2}} \left[ \Delta Q^c{}_{\gamma} , \Delta S^{\gamma}{}_c \right].
\ee
When we expand using the above coproducts, it can be shown the phase contributions cancel in the single site components of the Casimir.  As argued previously, these components give zero when acting on the states.  Just as important, if we examine the cross terms we retrieve the deformed Hamiltonian discussed in the previous section.  So, acting on a two-particle state $|\eta\rangle$, 
\be
 g_{AB}\Delta J^A \Delta J^B |\eta\rangle = \tilde{H}_{12} |\eta \rangle,
\ee
where $\tilde{H}_{12}$ is given by \eqref{defHam} if we 
%\be\begin{array}{rcl}
%\tilde{H}_2 &=& \left( c^{\dag}_a(1)c^{\dag}_b(2) - e^{i\alpha_{ab}} c^{\dag}_b(1)c^{\dag}_a(2)\right) c^b(2)c^a(1) \\[2mm] 
% & &+\left(c^{\dag}_a(1)a^{\dag}_{\alpha}(2) + e^{i\alpha_{a\alpha}} a^{\dag}_{\alpha}(1)c^{\dag}_a(2) \right) a^{\alpha}(2)c^a(1) \\[2mm]
% & &+\left(a^{\dag}_{\alpha}(1)c^{\dag}_a(2) + e^{i\alpha_{\alpha a}} c^{\dag}_a(1) a^{\dag}_{\alpha} (2)\right)c^a(2) a^{\alpha}(1) \\[2mm]
% & &+\left(a^{\dag}_{\alpha}(1)a^{\dag}_{\beta}(2) + e^{i\alpha_{\alpha\beta}} a^{\dag}_{\beta}(1)a^{\dag}_{\alpha}(2)\right) a^{\beta}(2) a^{\alpha}(1).
%\end{array}\ee
relate the deformation parameters with those from before, $\alpha_{IJ} = B_{IJ}$.  So, the deformed Hamiltonian commutes with all of the ordinary symmetry generators for arbitrary $B_{IJ}$,
\be\label{HJ}
 [ \tilde{H}_{12} , J_{12}\sp^A\sp_B] =0,
\ee
where we construct $J_{12}\sp^A\sp_B$ with the coproduct in \eqref{copr}.  An example two-site generator is, $R_{12}\sp^a\sp_b = R(1)^a\sp_b K(2)_{ba} + K(1)_{ab}R(2)^a\sp_b$.  With this information we can begin to reconstruct the identities associated with the Yangian structure given in the previous section.  To avoid confusion with the supercharge, $Q^a\sp_{\alpha}$, we shall denote the Yangian generator $Q^A$, in two index notation, as $\hat{J}^{A}\sp_{B}$.  Coproducts for twisted Yangian generators \cite{BasuMallick:1994pc, Bernard:1992ya} of \SU23 take the form
\be\label{deltaYang}\begin{array}{ccl}
\Delta \hat{R}^a\sp_b     %DELTA R
      &=& K_{ab}\otimes\hat{R}^a\sp_b + \hat{R}^a\sp_b\otimes K_{ba}
            + \frac{1}{2} h \left(R^a\sp_c K_{cb} \otimes K_{ca}R^c\sp_b - K_{ac}R^c\sp_b \otimes R^a\sp_c K_{bc}  \right) \\[2mm]
      & & +\frac{1}{2} h \left( Q^a\sp_{\gamma} K_{\gamma b} \otimes K_{\gamma a} S^{\gamma} \sp_b 
            +K_{a \gamma}S^{\gamma}\sp_b \otimes Q^a\sp_{\gamma} K_{b\gamma} \right) \\[2mm]
      & & -\frac{1}{6} h \delta^a_b \left( Q^c\sp_{\gamma} K_{\gamma c} \otimes K_{\gamma c} S^{\gamma} \sp_c 
            +K_{c \gamma}S^{\gamma}\sp_c \otimes Q^c\sp_{\gamma} K_{c \gamma} \right),\\[5mm]
\Delta \hat{L}^{\alpha}\sp_{\beta}    %DELTA L         
      &=& K_{\alpha\beta} \otimes \hat{L}^{\alpha}\sp_{\beta} + \hat{L}^{\alpha}\sp_{\beta}\otimes K_{\beta\alpha}
             +\frac{1}{2} h \left(L^{\alpha}\sp_{\gamma}K_{\gamma\beta} \otimes  K_{\gamma\alpha} L^{\gamma}\sp_{\beta} 
             - K_{\alpha\gamma}L^{\gamma}\sp_{\beta} \otimes L^{\alpha}\sp_{\gamma} K_{\beta\gamma} \right)\\[2mm]
      & & +\frac{1}{2} h \left( S^{\alpha}\sp K_{c\beta} \otimes K_{c\alpha} Q^c\sp_{\beta} 
             + K_{\alpha c}Q^c\sp_{\beta} \otimes S^{\alpha}\sp_c K_{\beta c} \right)\\[2mm]
      & & - \frac{1}{4} h \delta^{\alpha}_{\beta} \left(S^{\gamma}\sp_cK_{c\gamma}\otimes K_{c\gamma}Q^c\sp_{\gamma}
             +K_{\gamma c}Q^c\sp_{\gamma}\otimes S^{\gamma}\sp_c K_{\gamma c} \right),\\[5mm]
\Delta \hat{Q}^a\sp_{\alpha} %DELTA Q
      &=& K_{a\alpha}\otimes\hat{Q}^a\sp_{\alpha} + \hat{Q}^a\sp_{\alpha} \otimes K_{\alpha a}
             + \frac{1}{2} h \left(Q^a\sp_{\gamma}K_{\gamma \alpha} \otimes K_{\gamma a} L^{\gamma}\sp_{\alpha}
             -K_{a\gamma}L^{\gamma}\sp_{\alpha} \otimes Q^a\sp_{\gamma}K_{\alpha\gamma} \right)\\[2mm]
      & & + \frac{1}{2} h \left(R^a\sp_c K_{c\alpha}\otimes K_{ca}Q^c\sp_{\alpha} 
             - K_{ac}Q^c\sp_{\alpha} \otimes R^a\sp_c K_{\alpha c} \right),\\[5mm]
\Delta \hat{S}^{\alpha}\sp_a %DELTA S
      &=& K_{\alpha a}\otimes \hat{S}^{\alpha}\sp_a + \hat{S}^{\alpha}\sp_a \otimes K_{a\alpha}
             + \frac{1}{2} h \left(S^{\alpha}\sp_cK_{ca}\otimes K_{ca}R^c\sp_a
             - K_{\alpha c}R^c\sp_a \otimes S^{\alpha}\sp_cK_{ac} \right)\\[2mm]
      & & +\frac{1}{2} h \left(L^{\alpha}\sp_{\gamma} K_{\gamma a} \otimes K_{\gamma\alpha}S^{\gamma}\sp_a
             -K_{\alpha \gamma}S^{\gamma}\sp_a \otimes L^{\alpha}\sp_{\gamma}K_{a\gamma} \right),\\[5mm]
\Delta \hat{D} %DELTA D
      &=& 1\otimes \hat{D} + \hat{D}\otimes 1
             + \frac{1}{4}h \left(S^{\gamma}\sp_c K_{c\gamma} \otimes K_{c\gamma}Q^c\sp_{\gamma} 
             + K_{\gamma c}Q^c\sp_{\gamma}\otimes S^{\gamma}\sp_c K_{\gamma c} \right) .
\end{array}\ee
These coproducts are coassociative and quasi-cocommutative \cite{MacKay:2004tc}, and satisfy \eqref{alg}-\eqref{serre} in the double index basis.  In the derivation of the above coproduct for the deformed \SU23 Yangians, we had to respect the even/odd property of the generators and the traceless condition of the even generators.  The parameter $h$ is related to $\alpha$ in the Serre relation \eqref{serre}.
\par  We can use the twisted identity 
\be
\left[ \tilde{H}_{12} , q_{12}\sp^A\sp_B \right] = 8 h \hat{J}_{12}\sp^A\sp_B,
\ee
where $q_{12}\sp^A\sp_B =J^A\sp_B \otimes K_{BA} - K_{AB}\otimes J^A\sp_B $, and the $\hat{J}_{12}$ are given by the $h$ dependent terms in \eqref{deltaYang}.  For example, the two-site tree level Yangian generator $\hat{Q}^a\sp_{\alpha}$ is given by
\be\begin{array}{cl}
\hat{Q}_{12}\sp^{a}\sp_{\alpha} = & \frac{1}{2} h \left(Q(1)^a\sp_{\gamma}K(1)_{\gamma \alpha} K(2)_{\gamma a} L(2)^{\gamma}\sp_{\alpha}-K(1)_{a\gamma}L(1)^{\gamma}\sp_{\alpha}Q(2)^a\sp_{\gamma}K(2)_{\alpha\gamma} \right) \\[2mm]
&+ \frac{1}{2} h \left(R(1)^a\sp_c K(1)_{c\alpha}K(2)_{ca}Q(2)^c\sp_{\alpha} - K(1)_{ac}Q(1)^c\sp_{\alpha} R(2)^a\sp_c K(2)_{\alpha c} \right).
\end{array}\ee  Then on two-sites we can show
\be\label{HJhat}
\left[ \tilde{H}_{12} , \hat{J}_{12}\sp^A\sp_B \right] = {\textstyle \frac{1}{2}} h q_{12}\sp^A\sp_B,
\ee 
acting on all the eigenstates.  
\par In order to promote \eqref{HJ} and \eqref{HJhat} to L sites, we construct the L-site representation for $J^A\sp_B$ and $\hat{J}^A\sp_B$ using twisted coproducts with \eqref{copr} and \eqref{deltaYang}.  We find 
\be
\left[ \tilde{H} , J^A\sp_B \right] = 0,
\ee
and
\begin{align}\label{new}
\left[\tilde{H} , \hat{J}^A\sp_B \right] &= \sum^{L-1}_{i=1} \left[ \tilde{H}_{i,i+1} , \hat{J}_{i,i+1}\sp^A\sp_B \right] \cr
& =\frac{1}{2} h\left( J(1)^A\sp_B K(2)_{BA} \cdots K(L)_{BA} - K(1)_{AB}\cdots K(L-1)_{AB} J(L)^A\sp_B \right)\cr
\end{align}
If we examine an infinite length chain, which would resemble the worldsheet of the dual string theory, we can assume that surface terms at infinity can be dropped \cite{Dolan:2003uh}, and in that sense, $[ \tilde{H}, \hat{J}^A\sp_B ] =0$.  Thus, following the discussion in section 3, \eqref{new} provides a consistency check on the assumption that the \SU23 Yangian, with the twisted coproduct, holds to all orders in the Yang-Mills coupling constant.
\par  Up to this point, the analysis in this section holds for arbitrary, antisymmetric $\alpha_{IJ}$. We now illustrate the use of Yangians in these twisted theories in the two cases we examined earlier, in order to explain the residual symmetries.  
%CASE1
\vskip15pt
%\begin{subsection}{
\noindent {\em Case 1: $\gamma_1=\gamma_2=\gamma_3$}
\par We examine the twisted coproducts of Case 1.  Recall, the phase elements have the property $B_{a\alpha} =0$, $B_{\alpha\beta}=0$, and the $B_{ab}$ sector contains some non-zero entries.  We explicitly write the coproducts,
\be
K_{ab} = e^{\frac{i}{2}\left(\alpha_{ac} - \alpha_{bc}\right)E_{cc}} = K^{-1}_{ba},\hspace{3mm} K_{\alpha\beta} = 1 = K^{-1}_{\beta\alpha}, \hspace{3mm}K_{a \alpha} = e^{\frac{i}{2}\alpha_{ac}E_{cc}} = K^{-1}_{\alpha a}.
\ee
We examine the symmetry after using the twisted coproducts and find the residual SU(2)$\times$U(1)$^3$ symmetry corresponds to an undeformed coproduct:
\be\begin{array}{lcl}
\text{Residual Symmetry} & & \text{Remaining Symmetry}\\[2mm]
\Delta L^{\alpha}\sp_{\beta} =1 \otimes L^{\alpha}\sp_{\beta} + L^{\alpha}\sp_{\beta} \otimes 1, & \hspace{5mm} & \Delta R^{a}\sp_{b} = K_{ab} \otimes R^{a}\sp_{b} + R^{a}\sp_{b}\otimes K_{ba},\\[2mm]
\Delta D = 1 \otimes D + D \otimes 1, & &\Delta Q^{c}\sp_{\gamma} = K_{c\gamma} \otimes Q^{c}\sp_{\gamma} + Q^{c}\sp_{\gamma}\otimes K_{\gamma c},\\[2mm]
\Delta R^c\sp_c = 1\otimes R^c\sp_c + R^c\sp_c \otimes 1,& &\Delta S^{\gamma}\sp_{c}= K_{\gamma c} \otimes S^{\gamma}\sp_{c} + S^{\gamma}\sp_{c} \otimes K_{c \gamma}.
\end{array}\ee
Using these definitions one could check for the two-particle eigenstates, $\widetilde{|\pm\rangle}$ listed in a previous section,
\be
\left[ \tilde{H}_{12} , \hat{J}_{12}\sp^A\sp_B \right]\widetilde{|\pm\rangle} = {\textstyle \frac{1}{2}} h q_{12}\sp^A\sp_B \widetilde{|\pm\rangle}.
\ee
%Prior, we dealt with bare generators having no intrinsic state dependency.  It is for that reason the commutation relations of the Casimir failed to hold. Using the above definitions, one can check that all components of this twisted algebra do commute through the Casimir, and therefore with the Hamiltonian.  
%\end{subsection}
%CASE2
%\begin{subsection}
\vskip15pt
\noindent {\em Case 2: $\gamma_1=\gamma_2=-\gamma_3$}
\par We consider the richer structure of case 2.  From previous sections we saw a residual \su21$\times$U(1)$^2$ symmetry.  Recall, we have zero phase elements in the sectors $\alpha_{\alpha\beta} = \alpha_{3\alpha} =0$.  The other phases, $\alpha_{1\alpha}=-\alpha_{2\alpha}$ and $\alpha_{12}=\alpha_{13} = -\alpha_{23}$.  In this section, we label the fields $\Phi_I=\{\phi_a , \phi_3, \psi_{\alpha} \}$, with the indices $1 \leq a,b \leq 2$ and $1 \leq \alpha,\beta \leq 2$.  The twisted coproducts, have deformation parameters
\be\begin{array}{l}
K_{\alpha\beta} =1=K^{-1}_{\beta\alpha}, \hspace{3mm} K_{3\alpha} =1=K_{\alpha 3}, \hspace{3mm} K_{33}=1, \\[2mm]
K_{ab} = e^{\frac{i}{2}( \alpha_{a\gamma} - \alpha_{b\gamma})E_{\gamma\gamma} +\frac{i}{2}(\alpha_{a3}-\alpha_{b3})E_{33} + \frac{i}{2}(\alpha_{ac}- \alpha_{bc})E_{cc}} = K^{-1}_{ba}, \\[2mm]
K_{a3} = e^{\frac{i}{2}\alpha_{a\gamma}E_{\gamma\gamma}+\frac{i}{2}\alpha_{a3}E_{33} + \frac{i}{2}(\alpha_{ac} -\alpha_{3c})E_{33}} = K^{-1}_{3a}, \\[2mm]
K_{a\alpha} = e^{\frac{i}{2}\alpha_{a\gamma}E_{\gamma\gamma}+\frac{i}{2}(\alpha_{a3}-\alpha_{\alpha 3})E_{33}+\frac{i}{2}(\alpha_{ac}-\alpha_{\alpha c})E_{cc}} = K^{-1}_{\alpha a}.
\end{array}\ee
We apply these parameters and find the residual \su21$\times$U(1)$^2$ symmetry:
\be\begin{array}{lcl}
\text{Residual Symmetry} & & \text{Remaining Symmetry}\\[2mm]
\Delta L^{\alpha}\sp_{\beta} = 1 \otimes L^{\alpha}\sp_{\beta} + L^{\alpha}\sp_{\beta} \otimes 1, & & \Delta R^{a}\sp_{b} = K_{ab} \otimes R^{a}\sp_{b} + R^{a}\sp_{b}\otimes K_{ba},\\[2mm]
\Delta Q^{3}\sp_{\gamma}= 1\otimes Q^{3}\sp_{\gamma} + Q^{3}\sp_{\gamma}\otimes 1, & & \Delta Q^{c}\sp_{\gamma} = K_{c\gamma} \otimes Q^{c}\sp_{\gamma} + Q^{c}\sp_{\gamma}\otimes K_{\gamma c}, \\[2mm]
\Delta S^{\gamma}\sp_{3} = 1 \otimes S^{\gamma}\sp_{3} + S^{\gamma}\sp_{3} \otimes 1, & &\Delta S^{\gamma}\sp_{c}= K_{\gamma c} \otimes S^{\gamma}\sp_{c} + S^{\gamma}\sp_{c} \otimes K_{c \gamma}.\\[2mm]
\Delta D=1 \otimes D + D \otimes 1, \\[2mm]
\Delta R^c\sp_c = 1\otimes R^c\sp_c + R^c\sp_c \otimes 1,
\end{array}\ee
\vskip5pt
\noindent Again, one could directly compute, using the two-particle eigenstates in a previous section, to find $\left[ \tilde{H}_2 , \hat{J}^A\sp_B \right]\widetilde{|\pm\rangle} = q_{12}\sp^A\sp_B \widetilde{|\pm\rangle}$.
%\end{subsection}
\par In both cases,  since the coproducts for the remaining symmetries are non-standard and contain deformation parameters, these signal broken symmetries in the corresponding deformed gauge field theories.
\end{section}

%%%%%%%%%%%%%%%%%%%%%%%%%%%%%%%%%%%%
%%
%% SECTION: CONCLUSION
%%
%%%%%%%%%%%%%%%%%%%%%%%%%%%%%%%%%%%%
\begin{section}{Conclusion}
\par We identified the Yangian structure for \SU23.  Like the parent case of PSU$(2,2|4$), we could check that the Yangian commutation relations hold to one-loop.  Using the twisted R-matrix of the Yang-Baxter equation, supplied by Beisert and Roiban we computed the twisted coproducts using Reshetikhins formalism.  This twisted coproduct left a residual symmetry.  
\par  We derived that the useful identities found in the undeformed theory all have a twisted analog.  We explicitly calculated the twisted quadratic Casimir and showed, acting on two-particle states, the twisted Casimir is equivalent to the deformed Hamiltonian of the theory.  
\par We go on to show the residual symmetries found above have a null phase, $K_{AB}=1$,  corresponding to the standard, undeformed coproduct.  All remaining symmetries have phases associated with their coproducts.  So, in general to find a certain residual symmetry, one could start by assuming untwisted coproducts for the desired generators.
\par For chains of larger length, the twisted coproducts give a formalism that will maintain its `edge' effects and therefore can be used to check the Yangian symmetry extrapolated to one-loop.  Although higher loops in the \SU23 sector have dynamical lengths, our argument suggest this structure might survive to all orders, and therefore we expect to find also the \SU23 Yangian symmetry with twisted coproducts in the worldsheet of the dual string theory.  
\par Although the full \SU23 Yangian algebra is still present in the deformed theories, and is responsible for its integrability, the symmetries of the deformed field theories are the residual groups, due to the twisted coproducts.  These subgroups SU(2)$\times$U(1)$^3$ in Case 1, and \su21$\times$U(1)$^2$ in Case 2 correspond to the unbroken subgroups of SU$(2,2|1)\times$U(1)$^2$ in the \1N superconformal deformed gauge field theory, that survive in its \SU23 sector we consider in this paper.  
\par The twisted coproduct has provided a mechanism for maintaining integrability in a theory while breaking some of its initial symmetry.  This procedure might be useful in formulating integrable versions of even smaller symmetries, possibly ${\cal N} =0 $ Yang-Mills.  
\end{section}%CONCLUSION

%%%%%%%%%%%%%%%%%%%%%%%%%%%%%%%%%%%%%%%%%%%%%%%%%%%%%%%%%
%    ACKNOWLEDGMENTS
\acknowledgments{
I would like to thank Louise Dolan for her many conversations.  This work was partially supported by the U.S. Department of Energy, Grant No. DE-FG01-06ER06-01, Task A.}
%%%%%%%%%%%%%%%%%%%%%%%%%%%%%%%%%%%%%%%%%%%%%%%%%%%%%%%%%%%%%%
%%%%%%%%%%%%%%%%%%%%%%%%%%%%%%%%%%%%%%%%%%%%%%%%%%%%%%%%%%%

%%%%%%%%%%%%%%%%%%%%%%%%%%%%%%%%%%%%
%%
%% APPENDIX
%%
%%%%%%%%%%%%%%%%%%%%%%%%%%%%%%%%%%%%
\begin{appendix}

%%%
% A: BASIS (SINGLE)
%%%
\begin{section}{Single Index Basis}
\par It is sometimes convenient to chose a single index basis for the twenty-four generators of \SU23.  The twelve even generators have the representation
\begin{equation}\label{evenBasis}
\begin{array}{lllll}
\text{{\bf SU(3)}}& & \text{{\bf SU(2)}} & & \text{{\bf U(1)}}\\[2mm]
J^1 = R^1\sp_2+R^2\sp_1 & &  J^9=L^1\sp_2+L^2\sp_1 & &J^{12}=D \\[2mm]
J^2 = i(R^1\sp_2-R^2\sp_1) & & J^{10} = i(L^1\sp_2-L^2\sp_1) \\[2mm]
J^3 = R^1\sp_1 -R^2\sp_2 & & J^{11} = L^1\sp_1 - L^2\sp_2 \\[2mm]
J^4 = R^1\sp_3+R^3\sp_1 \\[2mm]
J^5 = i(R^1\sp_3-R^3\sp_1) \\[2mm]
J^6 = R^2\sp_3+R^3\sp_2 \\[2mm]
J^7 = i(R^2\sp_3 - R^3\sp_2) \\[2mm]
J^8 = R^1\sp_1+R^2\sp_2 - 2R^3\sp_3
\end{array}
\end{equation}
The twelve odd generators have the representation
\begin{equation}\label{oddBasis}
\begin{array}{lll}
J^{13} = S^1\sp_1 + Q^1\sp_1    & & J^{19} = S^2\sp_1 + Q^1\sp_2\\[2mm]
J^{14} = i(S^1\sp_1 - Q^1\sp_1) & & J^{20} = i(S^2\sp_1 + Q^1\sp_2)\\[2mm]
J^{15} = S^1\sp_2 + Q^2\sp_1    & & J^{21} = S^2\sp_2 + Q^2\sp_2\\[2mm]
J^{16} = i(S^1\sp_2 - Q^2\sp_1) & & J^{22} = i(S^2\sp_2 + Q^2\sp_2)\\[2mm]
J^{17} = S^1\sp_3 + Q^3\sp_1    & & J^{23} = S^2\sp_3 + Q^3\sp_2\\[2mm]
J^{18} = i(S^1\sp_3 - Q^3\sp_1) & & J^{24} = i(S^2\sp_3 + Q^3\sp_2).
\end{array}
\end{equation}
  The metric is symmetric (and diagonal) in the even regions and antisymmetric in the odd regions.
\begin{equation}\label{metric}\begin{array}{c}
g_{(1)(1)} = g_{(2)(2)} = g_{(3)(3)} = g_{(4)(4)} =g_{(5)(5)} =g_{(6)(6)} =g_{(7)(7)} = g_{(8)(8)} = -\frac{1}{4}, \\[2mm]
g_{(9)(9)} = g_{(10)(10)} = g_{(11)(11)} = \frac{1}{4}, \\[2mm]
-g_{(13)(14)} = g_{(14)(13)} = -g_{(15)(16)} = g_{(16)(15)} =\cdots =g_{(22)(21)}= -g_{(23)(24)} = g_{(24)(23)} = \frac{1}{4i}.
\end{array}\end{equation}
The \SU23 algebra obeys the commutation $[J^A,J^B] = f^{ABC}g_{CD}J^D$.  The structure constants are totally antisymmetric with an additional minus sign under the interchange of two odd indices.
\begin{equation}
\begin{array}{l}
f^{(1)(2)(3)} = -f^{(9)(10)(11)}=-8i \\[2mm]
f^{(1)(4)(7)} = f^{(1)(6)(5)} = f^{(2)(4)(7)} = f^{(2)(5)(7)} = f^{(3)(4)(5)}=f^{(3)(7)(6)}= -4i \\[2mm]
f^{(4)(5)(8)}=f^{(6)(7)(8)}= -12i \\[2mm]
f^{(1)(13)(15)}=f^{(1)(14)(16)}=f^{(1)(19)(21)}=f^{(1)(20)(22)}=-4\\[2mm]
f^{(2)(13)(16)} = - f^{(2)(14)(15)} = f^{(2)(19)(22)} = -f^{(2)(20)(21)} = -4\\[2mm]
f^{(3)(13)(13)}=f^{(3)(14)(14)}=-f^{(3)(15)(15)} = -f^{(3)(16)(16)}=-4\\[2mm]
f^{(3)(19)(19)}=f^{(3)(20)(20)}=-f^{(3)(21)(21)} = -f^{(3)(22)(22)}=-4\\[2mm]
f^{(4)(13)(17)} = f^{(4)(14)(18)}=f^{(4)(19)(23)}=f^{(4)(20)(24)} = -4\\[2mm]
f^{(5)(13)(18)} = f^{(5)(14)(17)}=f^{(5)(19)(24)}=f^{(5)(20)(23)} = -4\\[2mm]
f^{(6)(15)(17)} = f^{(6)(16)(18)}=f^{(6)(21)(23)}=f^{(6)(22)(24)} = -4\\[2mm]
f^{(7)(15)(18)} = -f^{(7)(16)(17)}=f^{(7)(21)(24)}=-f^{(7)(22)(23)} = -4\\[2mm]
f^{(8)(13)(13)} = f^{(8)(14)(14)}=f^{(8)(15)(15)}=f^{(8)(16)(16)} = -4 \\[2mm]
f^{(8)(19)(19)} = f^{(8)(20)(20)}=f^{(8)(21)(21)}=f^{(8)(22)(22)} = -4\\[2mm]
f^{(8)(17)(17)} = f^{(8)(18)(18)}=f^{(8)(23)(23)}=f^{(8)(24)(24)} = 8\\[2mm]
f^{(9)(13)(19)}=f^{(9)(14)(20)}=f^{(9)(15)(21)}=f^{(9)(16)(22)}=f^{(9)(17)(23)}=f^{(9)(18)(24)} = 4\\[2mm]
f^{(10)(13)(20)}=-f^{(10)(14)(19)}=f^{(10)(15)(22)}=-f^{(10)(16)(21)}=f^{(10)(17)(24)}=-f^{(10)(18)(23)} = -4\\[2mm]
f^{(11)(13)(13)} = f^{(11)(14)(14)} =f^{(11)(15)(15)} =f^{(11)(16)(16)} = f^{(11)(17)(17)} = f^{(11)(18)(18)} =4\\[2mm]
f^{(11)(19)(19)} = f^{(11)(20)(20)} =f^{(11)(21)(21)} =f^{(11)(22)(22)} = f^{(11)(23)(23)} = f^{(11)(24)(24)} =-4\\[2mm]
f^{(12)(13)(13)} = f^{(12)(14)(14)}=f^{(12)(15)(15)}=f^{(12)(16)(16)}=f^{(12)(17)(17)}=f^{(12)(18)(18)}=2\\[2mm]
f^{(12)(19)(19)} = f^{(12)(20)(20)}=f^{(12)(21)(21)}=f^{(12)(22)(22)}=f^{(12)(23)(23)}=f^{(12)(24)(24)}=2.
\end{array}
\end{equation}
\end{section}%BASIS

%%%
% SECTION: TWO INDEX BASIS
%%%
\begin{section}{Double Index Basis}
The generators can be written in a double index notation \cite{Bernard:1992ya,BasuMallick:1994pc}.  For a general superalgebra, we can define matrices $\left(E_{AB}\right)_{ij} = \delta_{Ai}\delta_{Bj}$ which satisfy $\left[E_{AB} , E_{CD} \right] = \delta_{CB}E_{AD} - \delta_{AD}E_{CB}$.  For \SU23, 
\be\begin{array}{l}
\left[ R^a\sp_b , R^c\sp_d \right] = \delta^a_d R^c\sp_b - \delta^c_b R^a\sp_d, \hspace{5mm} \left[R^a\sp_b , Q^c\sp_{\gamma}\right] = -\delta^c_bQ^a\sp_{\gamma} + \frac{1}{3}\delta^a_bQ^c\sp_{\gamma} , \hspace{5mm} \left[R^a\sp_b , S^{\gamma}\sp_c \right] =  \delta^a_cS^{\gamma}\sp_b - \frac{1}{3} \delta^a_b S^{\gamma}\sp_c, \\[2mm]

\left[ L^{\alpha}\sp_{\beta} , L^{\gamma}\sp_{\delta}\right] = \delta^{\alpha}_{\delta}L^{\gamma}\sp_{\beta} - \delta^{\gamma}\sp_{\beta}L^{\alpha}\sp_{\delta}, \hspace{5mm} \left[ L^{\alpha}\sp_{\beta} , Q^c\sp_{\gamma} \right] = \delta^{\alpha}_{\gamma}Q^c_{\beta} - \frac{1}{2} \delta^{\alpha}_{\beta} Q^c\sp_{\gamma}, \hspace{5mm} \left[L^{\alpha}\sp_{\beta} , S^{\gamma}\sp_c \right] = -\delta^{\gamma}_{\beta} S^{\alpha}_c + \frac{1}{2} \delta^{\alpha}_{\beta} S^{\gamma}_c, \\[2mm]

\left\{ Q^a\sp_{\alpha} , S^{\beta}\sp_b \right\} = \delta_{\alpha}^{\beta} R^a\sp_b + \delta^{a}_{b} L^{\beta}\sp_b + \frac{1}{3}\delta^a_b\delta^{\beta}_{\alpha}  D, \hspace{5mm} \left[ D , Q^a\sp_{\alpha}\right] = +\frac{1}{2} Q^a\sp_{\alpha}, \hspace{5mm} \left[ D, S^{\alpha}\sp_a \right] = - \frac{1}{2} S^{\alpha}_a.
\end{array}\ee
We transform between our \SU23 generators and the general matrices defined above by
\be\begin{array}{c}
R^a\sp_b =  E_{ba} - \frac{1}{3}\delta^a_b E_{cc}, \hspace{5mm} L^{\alpha}\sp_{\beta} = E_{\beta\alpha} - \frac{1}{2} \delta^{\alpha}_{\beta}E_{\gamma\gamma}, \\[2mm]
S^{\gamma}\sp_c =E_{c\gamma}, \hspace{5mm} Q^c\sp_{\gamma} = E_{\gamma c},\hspace{5mm} D= E_{cc} + \frac{3}{2}E_{\gamma\gamma},
\end{array}\ee
where repeated indices are summed over.
\end{section}
\end{appendix}

%%%%%%%%%%%%%%%%%%%%%%%%%%%%%%%%%%%%%%%%%%%%%%%%%%%%%%%%%%%%%%%
%%%%%%%%%%%%%%%%%%%%%%%%%%%%%%%%%%%%%%%%%%%%%%%%%%%%%%%%%%%%%%%

\end{document}